\documentclass[12pt]{JHEP3}
\usepackage{bm}
\usepackage{amssymb,amsmath,amsthm}
\usepackage{mathrsfs} %script fonts 
\RequirePackage{graphicx}

\newcommand{\be}{\begin{equation}} 
\newcommand{\ee}{\end{equation}}
\newcommand{\bea}{\begin{eqnarray}}
\newcommand{\eea}{\end{eqnarray}}

\newcommand{\gapp}{\mathrel{\raise.3ex\hbox{$>$}\mkern-14mu
              \lower0.6ex\hbox{$\sim$}}}
\newcommand{\lapp}{\mathrel{\raise.3ex\hbox{$<$}\mkern-14mu
              \lower0.6ex\hbox{$\sim$}}}

\newcommand\lsim{\lesssim}
\newcommand\gsim{\gtrsim}

\newcommand\vev[1]{{\langle {#1} \rangle}}
\renewcommand\({\left(}
\renewcommand\){\right)}
\renewcommand\[{\left[}
\renewcommand\]{\right]}

\newcommand\eq[1]{Eq.~(\ref{#1})}
\newcommand\eqs[2]{Eqs.~(\ref{#1}) and (\ref{#2})}
\newcommand\eqss[3]{Eqs.~(\ref{#1}), (\ref{#2}), and (\ref{#3})}

\newcommand\eqreff[1]{(\ref{#1})}
\newcommand\eqsref[2]{(\ref{#1}) and (\ref{#2})}

\newcommand\pa{\partial}

\newcommand\mpl{M_{\rm P}}

\newcommand{\dlabel}[1]{\label{#1}}

\def\calp{{\cal P}}

\def\calpz{{\calp_\zeta}}

\newcommand\bfk{{\mathbf k}}

\newcommand\bfx{{\mathbf x}}

\newcommand\GeV{\,\mbox{GeV}}
\newcommand\MeV{\,\mbox{MeV}}

\newcommand\sub[1]{_{\rm #1}}

\newcommand\mone{^{-1}}
\newcommand\mtwo{^{-2}}
\newcommand\mthree{^{-3}}
\newcommand\mfour{^{-4}}
\newcommand\mfive{^{-5}}
\newcommand\mhalf{^{-1/2}}
\newcommand\half{^{1/2}}
\newcommand\third{^{1/3}}

\newcommand\twothird{^{2/3}}

\newcommand\quarter{^{1/4}}

\newcommand\threehalf{^{3/2}}
\newcommand\mthreehalf{^{-3/2}}

\newcommand{\fnl}{f\sub{NL}}

\newcommand{\calpzphi}{\calp_{\zeta_\phi}}

\newcommand{\calpzlin}{\calp_{\zeta\slin}}

\newcommand{\start}{\sub{start}}
\newcommand{\send}{\sub{end}}
\newcommand{\snl}{\sub{nl}}
\newcommand{\slin}{\sub{lin}}
\newcommand{\smax}{_=}

\newcommand\modm[1]{|m({#1})|}
\newcommand{\mmt}{\modm{t}}

\newcommand{\mmts}{|m^2(t)|}
\newcommand{\kpeak}{k\sub{peak}}

\newcommand{\chivev}{\chi\sub{vev}}

\title {The hybrid inflation waterfall and  the primordial         
curvature perturbation\footnote                                  
{A preliminary version of this paper appeared as arXiv:1107.1681}}

\author{David H. Lyth,
Consortium for Fundamental Physics, Cosmology and Astroparticle Group, 
Department of Physics, Lancaster University, 
Lancaster LA1 4YB, UK}

\abstract{  
Without demanding a specific form for the inflaton 
 potential,  we obtain an estimate of the
contribution  to the curvature perturbation
 generated during the linear era of the hybrid
inflation waterfall.  
The  spectrum  of this contribution  peaks at some  wavenumber $k=k_*$, and
 goes like $k^3$ for $k\ll k_*$, making it typically negligible on
 cosmological scales. The scale
 $k_*$ can be outside the horizon at the end of inflation, in which case
$\zeta=- (g^2 - \vev{g^2})$ with $g$ gaussian. Taking this into account,
 the  cosmological bound on the abundance of black holes is likely to be
 satisfied if the  curvaton
mass $m$ much bigger than the Hubble parameter $H$, but is likely to 
be violated if
 $m\lsim H$. Coming to  the contribution  to $\zeta$
from the rest of the waterfall, we  are led to consider
the use of the `end-of-inflation' formula, giving the contribution
to $\zeta$  generated during a sufficiently 
sharp transition from nearly-exponential
inflation to  non-inflation, and we state
 for the first time the criterion for the transition
to be sufficiently sharp. 
 Our formulas are applied to
 supersymmetric GUT inflation and to supernatural/running-mass inflation.}

\preprint{}

\begin{document}

\section{Hybrid inflation}

\dlabel{shyb}

Hybrid inflation \cite{earlyhybrid,andreihybrid,ourhybrid} 
 ends with a phase transition known as the waterfall, which up to now has been
studied only in special cases. 
This  paper, which is a continuation of \cite{p10},  provides
 a  rather general  treatment. We begin by defining the setup.

\subsection{Scales leaving the horizon}

An  inflation model starts to make contact with observation only
around the time that  the observable universe leaves the horizon.
The following description hybrid inflation is intended to apply to the 
subsequent era.

  Within the standard cosmology, 
the number  $N\sub{obs}$ of $e$-folds  of inflation 
after the observable universe leaves the horizon 
satisfies \cite{book}
\be
63-\frac12 \ln \frac{10\mfive\mpl}{H} \lsim  N\sub{obs} \lsim 
 49 - \frac13 \ln \frac{10\mfive\mpl}{H}
. \dlabel{nobs} \ee
(The time-dependence of $H$ is ignored in this expression, which
is usually a good approximation.)
The upper bound corresponds to
matter domination from the end of inflation to the epoch $T= 1\MeV$,
with radiation domination thereafter until the  observed matter dominated era,
while the lower bound replaces the former era by one of radiation domination.

The scales probed  by observation of large scale structure
(cosmological scales) leave the horizon during the first 15 or so $e$-folds
after the observable universe. On these scales, the curvature perturbation
$\zeta$ is nearly gaussian with a  a nearly scale-invariant spectrum
 $\calpz(k)\sim (5\times 10\mfive)^2$. 

\subsection{Hybrid inflation potential}

\dlabel{sshybrid}

Our analysis applies to a wide class of hybrid inflation models. The
essential features of the potential are captured by the following expression;
\bea
V(\phi,\chi) &=& V_0 + V(\phi)
+\frac12 m^2(\phi)  \chi^2 + \frac14\lambda \chi^4
 \dlabel{fullpot}  \\
m^2(\phi) &\equiv & g^2\phi^2  -m^2 \equiv g^2 \( \phi^2-\phi\sub c^2 \)
\dlabel{mmtsdef} . \eea
To  have a perturbative quantum theory we demand
$g\ll 1$ and $\lambda\ll 1$.
The    inflaton $\phi$ is supposed to have  zero vev and $V(\phi)$ 
is set  to zero at the vev. We require $V'(\phi)>0$ during inflation so that
$\phi$ moves towards its vev. The era of inflation with $\phi<\phi\sub c$
is called the waterfall.

The requirements that $V$ and $\pa V/\pa \chi$ vanish in the vacuum 
determine $V_0$ and  the vev of the waterfall field $\chi$:
\be
\chivev^2 =m^2/\lambda,\qquad V_0=m^4/4\lambda 
. \dlabel{vevs} \ee

It is necessary for our analysis that the waterfall field $\chi$ has the 
canonical kinetic term. 
For simplicity we will pretend that $\chi$ is a single real field. At least within
the Standard Scenario defined below, that 
cannot really be the case because it would lead to 
 the formation of  domain walls,  located along surfaces at which
$\chi(\bfx,t)$ is trapped at the origin, which 
would be fatal to the cosmology. In reality, $\chi$ will be replaced by
 a function of two or  more real fields. So that there is only one effective
degree of freedom in the $\chi$ direction, we will demand that the function 
is invariant under some symmetry group of the action.
Then  the only change in our analysis for the realistic case 
would be the introduction of some numerical factors into the equations.
In the realistic case  the domain walls might be replaced by 
  cosmic strings  or monopoles, 
but in general  the trapping of $\chi$ will not occur and $\chi(\bfx,t)$ will 
everywhere  approach its vev.

The inflaton $\phi$ may also be replaced by a function of two or more real
fields. If there is still only one effective degree of freedom the only
change is again  the introduction of numerical factors. In the opposite
case of   multi-field inflation,  corresponding to  a family of  inflationary
trajectories that are curved in field space, most of 
our analysis still applies if, by the onset of the waterfall,
the family has collapsed to a single  effective trajectory which has negligible
curvature during the waterfall.  To obtain powerful results
 we take the inflaton
to have the  canonical kinetic term, though much of our analysis would apply
to, for instance, k-inflation \cite{kinflation}.

Hybrid inflation was first discovered  in 
 in the context of single-field inflation  \cite{earlyhybrid,andreihybrid}.
It was given its name in 
\cite{andreihybrid}, where the form \eqreff{fullpot} was invoked for 
$V(\phi,\chi)$ with  $V(\phi)=m_\phi^2 \phi^2/2$.
With parameters chosen to give the Standard  Scenario,
and demanding also that 
$\delta\phi$ is responsible for the observed
curvature perturbation,  this 
 gives spectral index $n>1$ in contradiction with observation.
 Many forms of $V(\phi)$ have been proposed, which
allow $\delta\phi$ to generate the curvature 
perturbation \cite{al,book}  within the single-field inflation scenario. 

In our calculations we employ  \eq{fullpot}
for $V(\phi,\chi)$, without specifying the inflaton potential $V(\phi)$.
Minor variants of \eq{fullpot} would make little difference.
The  interaction $g^2\phi^2\chi^2$ might be replaced by
$\phi^2 \chi^{2+n}/\Lambda^n$ where $\Lambda$ is a uv cutoff,
or  the  term $\lambda\chi^4$ might be replaced by $\chi^{4+n}/\Lambda^n$.
 For our purpose, these  variants are  equivalent to allowing
(respectively) $g$ and $\lambda$ to be  many orders of magnitude below
unity. 

More drastic modifications of \eq{fullpot} have been proposed, 
 including inverted hybrid inflation \cite{inverted}
where $\phi$ is increasing during inflation, as well as  mutated and smooth hybrid
inflation \cite{mutated,smooth}
where the waterfall field varies during inflation. Also, the waterfall potential
might have a local minimum at the origin so that the waterfall proceeds by
bubble formation \cite{firstorder,ourhybrid}. 
Our analysis does not apply to those cases.

\subsection{Standard Scenario}

\dlabel{sstandard}

By varying the parameters in the potential \eqreff{fullpot},  one can have
 a wide range of scenarios that is still not fully explored.
Most discussions of hybrid inflation make some assumptions, corresponding
to what might be called the Standard Scenario. 
In this section we state those assumptions, which are made
in the rest of the paper. 

Until $\chi$ approaches its vev at the end of the waterfall,
inflation is supposed to be nearly exponential 
($\epsilon_H \equiv |\dot H|/H^2 \ll 1$) with 
 $V_0$ dominating  the potential:
\be
3\mpl^2 H^2(t) = \rho(t) \simeq V_0
. \dlabel{hofv} \ee
We take $H$ to be constant during the 
waterfall, which is usually a good approximation.
Nearly exponential  inflation requires 
\be
\dot\phi^2 \ll 3\mpl^2 H^2
.\dlabel{dotphibound} \ee

\eqs{vevs}{hofv}  give
\be
\chivev^2/\mpl^2\simeq  12 H^2/m^2
 \dlabel{chiandv}. \ee
It  is usually supposed that  $\chivev\ll \mpl$ corresponding to $m\gg H$.
(In particular,
GUT inflation \cite{gut1,gut2}
takes $\chi$ to be a GUT  Higgs field 
with $\chivev\sim 10\mtwo\mpl$.)
One sometimes considers $\chivev$ roughly of order
$\mpl$ corresponding to $m$  roughly of order $H$
(supernatural \cite{supernatural} and running mass 
\cite{running} inflation). 
There do not seem to be any papers considering
 $\chivev\gg \mpl$ which would correspond to $m\ll H$.

Using \eqs{vevs}{hofv}, the requirement $\lambda\ll 1$ is equivalent to
\be
m/H <  \sqrt{\mpl/H}\qquad (\lambda \ll 1)
. \ee
Successful BBN and the upper bound on the tensor perturbation require
\cite{book}
\be
10^{-42} < H/\mpl < 10\mfive \qquad \mbox{(BBN and tensor)}
. \dlabel{hbound} \ee
The upper part of the range is favoured, especially \cite{mytev}
because we deal with hybrid inflation.

One usually requires $\phi\ll \mpl$ 
but we will just invoke the weaker requirement\footnote
{This is also invoked in our earlier paper \cite{p10}
but note that Eq.~(80) of \cite{p10} has a typo.}
\be
\phi\sub c\equiv m/g \ll \mpl
\dlabel{phibound} . \ee

If $\phi$ is big enough we have $m^2(\phi) \gsim H^2$. 
Then  we assume that
 $\chi$ vanishes  up to a    vacuum fluctuation which is set
 to zero. 
If $\phi$  is  small enough, $m^2(\phi) \lsim -H^2$.
 Hence there is a `transition'  regime with 
 $|m^2(\phi)| \ll  H^2$. If the transition  takes several
 Hubble times,  the quantum
fluctuation of $\chi$ will be converted to a 
classical perturbation, with spectrum
$\sim (H/2\pi)^2$ on all scales leaving the horizon during the transition. 
To avoid this the  transition should take less than a Hubble time
or so (fast transition).

The waterfall starts at $m^2(\phi) =0$ which is in the middle of the
transition. During the waterfall
the vacuum fluctuation of $\chi$ is converted to 
a classical field $\chi(\bfx,t)$, with   $\chi^2$  moving  towards $\chivev^2$.
The waterfall ends when $\chi^2(\bfx,t)\simeq \chivev^2$,
and inflation is  supposed to end then
 because $V(\phi)$ is not 
supposed to support inflation without the additional
term $V_0$. 

Regarding $\phi$, we require that it decreases monotonically before the waterfall,
and afterward for  as long as it affects the evolution of $\chi$.
This   assumption   is not at all trivial, 
because $V(\phi)$ may steepen as $\phi$ decreases, causing $\phi$ to
 oscillate   about the origin.
The  evolution of $\chi$  has yet to be studied for that case,  which occurs
in part of the parameter space for 
some well-motivated forms of the potential, including GUT inflation \cite{gut1,gut2}
and  running-mass inflation \cite{running}.

\subsection{The waterfall}

\dlabel{swater}

During the waterfall we need to consider both $\phi$ and $\chi$.
Taking both fields to live in unperturbed spacetime (ie.\ ignoring 
back-reaction)  the evolution equations during the waterfall are
\bea
\ddot   \phi + 3H\dot \phi  +    V'(\phi)  -\nabla^2  \phi
& =&  - g^2\chi^2 \phi \dlabel{fullphi} \\
\ddot \chi + 3 H\dot\chi +  m^2(\phi)\chi -\nabla^2\chi  &=&   
 - \lambda \chi^3 \dlabel{fullchi} .
\eea 

We assume that the waterfall starts with an era during which \eq{fullchi} 
 can be replaced by\footnote
{This assumption   implies  some 
 lower bound on $|\dot\phi|$ but it is  not clear how to calculate the  bound
\cite{p10,dufaux}.}
\be
\ddot \chi_\bfk + 3H \dot\chi_\bfk + \[ (k/a)^2 + m^2(t)) \] \chi_\bfk
= 0
, \dlabel{chiddot} \ee
with  $m^2(t)$ is independent of $\chi$. 
We call this the linear era, and it will be our
 main focus.
Regarding $\chi_\bfk(t)$ as an operator \cite{p10,book}, 
its mode function $\chi_k(t)$
also satisfies \eq{chiddot}. We  will see how $\chi_k(t)$
grows exponentially for suitably small $k$, generating a classical quantity
$\chi_\bfk(t)$.
Keeping only the classical modes, we arrive at a classical field
$\chi$.

During at least the first part of the linear era,
$m^2(\phi)$ depends significantly on $\phi$. Then the 
right hand side  of \eq{fullphi}  has to  be negligible 
so that $m^2(t)$ can be independent of $\chi$.
With that condition in place  we just have to worry about
the perturbation $\delta\phi$ that is generated from the vacuum fluctuation.

If the linear era of the 
waterfall takes no more than a Hubble time or so, $\delta\phi$ can
be completely 
 eliminated by taking the spacetime slicing to be one of uniform $\phi$.
But if  the waterfall takes much more than a Hubble time,
 new contributions to $\delta\phi$ are generated
as each scale leaves the horizon. To avoid this quantum effect
on the evolution $\chi$, we have to 
 assume that the new contributions to
$\delta\phi$ are  negligible. Then we can again choose the slicing
of uniform $\phi$.\footnote
{Since we neglect the new contributions to  $\delta\phi$ during
the linear era (while $m^2(\phi)$ depends significantly on $\phi$),
 we  neglect also
their effect on  the spectrum $\calpz(k)$.  That is presumably a good approximation
if $\calpzphi$ given by \eq{calpzphi} is much less than the contribution
$\calp_{\zeta\slin}(k)$ that we are going to calculate. That will probably be the case
if $\calp_{\zeta\slin}(k)$ is big enough to form black holes, which is our main 
concern. It will not be the case if
 we deal with one of those exceptional inflation models where $\calpzphi(k)$,
on the scales leaving the horizon during the linear era of the waterfall,
is itself big enough to form black holes. Then we have the opposite situation:
\eq{calpzphi} will be valid if $\calp_{\zeta\slin}$ is {\em not} big enough to
form black holes.}

For the threading of spacetime, we choose the
 comoving worldlines (those moving with the fluid, so that a fluid element has zero
momentum density). The  gradients  of both $\phi$ and (as we shall see) the classical field
 $\chi$  are  small  compared with their time derivatives. If they vanished, the comoving
worldlines would be  free-falling and orthogonal to the 
slicing, and we could choose the time coordinate labeling the slicing
to be proper time along each thread. We assume that the gradients are small enough
to make that choice possible to an acceptable approximation.
This completes  the definition of the gauge in which
the classical field  $\chi(\bfx,t)$ is defined.

We also need to justify the use of \eq{chiddot} for the mode function
$\chi_k$,
before the
classical quantity $\chi_\bfk(t)$  is generated. 
As we will see, \eq{chiddot} is needed for that purpose only 
for  modes that are well inside the horizon during this time and
(at least with the approximation of Section \ref{sending}) 
 only for much less than a Hubble time. That being the case,
we  can  ignore the second term of \eq{chiddot} 
and set $a$ equal to a constant so that \eq{chiddot} becomes a flat
spacetime equation in which back-reaction is negligible.

\section{Waterfall field during the linear era}

\dlabel{smggh} 

\subsection{Evolution of $\chi$}

With $H$ constant,  we can use
 conformal time  $\eta=-1/aH$ to write  \eqreff{chiddot} as
\be
\frac{d^2(a\chi_\bfk)}{d\eta^2} +\omega_k^2 a\chi_\bfk =0,  \dlabel{fulleq} 
\ee
with 
\be
\omega_k^2(\eta) \equiv  k^2 +a^2\tilde m^2(t),\qquad \tilde m^2\equiv  
m^2(t)-2H^2,\qquad m^2(t) \equiv g^2\phi^2(t) - m^2
 .  \ee

For sufficiently small $k$, 
we can set $\omega_k^2\simeq  \omega_{k=0}^2=a^2\tilde m^2$.
Then $\omega_k^2$ switches  from positive to negative before  
$\phi=\phi\sub c$ (but much less than a Hubble time before, by 
 virtue of our fast transition assumption).
 For $k^2>0$ the switch is later.
For the scales that we need to consider, we   assume that 
 there  are eras  both before and after the switch 
when  $\omega_k^2$ satisfies the adiabaticity condition
$d|\omega_k|/d\eta  \ll |\omega_k^2|$. 

Taking $\chi_\bfk$ to be an operator, its mode function $\chi_k$ satisfies
\eq{fulleq}.
During the adiabaticity era before the switch we take the mode function to be
\be
a\chi_k =  (2\omega_k(\eta))\mhalf \exp\( -i\int^\eta
 \omega_k(\eta) d\eta \)
, \dlabel{before} \ee
which defines the vacuum state.
During the adiabaticity era after the switch 
\be
a\chi_k \simeq  (2|\omega_k(\eta)|)\mhalf \exp\( \int^\eta_{\eta_1(k)}
 |\omega_k(\eta)| d\eta \) \dlabel{after}
, \ee
where the subscript 1 denotes the beginning of the adiabatic era.
The displayed prefactor is exact  \cite{p10} only  
if 
$m^2(t)\propto t$ and $H(t-t_1)\ll 1$. As we are about to see, $\chi_k$ grows during 
this era and we call it the growth era.

During  the growth era,
 the adiabaticity condition 
is equivalent to the two   conditions
\bea
\frac{2H}{\mmt}  &\ll&
\[ 1 - \(\frac k{a(t) \mmt } \)^2 \]^{3/2} \dlabel{adcon2} \\
 \frac1{\mmt^2}\frac{d\mmt}{dt}
 &\ll&
\[ 1 - \(\frac k{a(t) \mmt } \)^2 \]^{3/2}
. \dlabel{adcon3} \eea
The growth  era begins when both conditions are first satisfied. 

The first condition implies $\mmt\gg H$ so that $|\tilde m(t)|  \simeq \mmt$,
and it can hold only if $m\gg H$. For $k\ll a(t) \mmt$ we have
$|\omega_k| \simeq  |\omega_{k=0}| \simeq  a(t)\mmt$. 
With  \eq{after} this gives $\dot \chi_k \simeq |\omega_k|\chi_k/a$.
At $k=0$ \eq{after} becomes
\be
 \chi_{k=0}(t)  \simeq (2a^3\mmt)\mhalf
\exp\( \int^t_{t_1} dt \mmt   \) 
. \dlabel{chik1} \ee
Ignoring the relatively slow time-dependence of the prefactor, we 
have as a rough approximation
\be
 \chi_{k=0}(t)  \simeq (2a_1^3|m(t_1)|)\mhalf
\exp\( \int^t_{t_1} dt \mmt   \) 
\dlabel{chik0} . \ee

In the regime $k\ll a\mmt$ we have
\be |\omega_k| \simeq a\mmt \(1- \frac12 \frac{k^2}{a^2\mmts} \)
, \ee
giving 
\be
 \chi_k(t)  \simeq  \chi_{k=0}(t)  e^{-k^2/2k_*^2(t)}, \dlabel{chik} \ee
where 
\be
k_*^2(t) \equiv\(  \int^t_{t_1}\frac{dt}{a^2\mmt} \)\mone
. \dlabel{kstarexp} \ee

By virtue of \eq{adcon3}, the change in $\mmt$ in time $\mmt\mone$ is 
small and so is the change in $a$. Defining  
\be
t\start\equiv t_1+\modm{t_1}\mone
, \dlabel{tstart} \ee
 we get 
\be
k_*(t\start)\simeq a(t_1)\modm{t_1}
. \dlabel{kstart} \ee
To avoid the divergence in $k_*(t)$ at $t=t_1$, 
we will regard $t\start$  as the start of the growth era
rather than $t_1$. After $t\start$,
 $k_*(t)$ decreases while $a\mmt$ increases. We assume that
$k_*^2(t)\ll  a^2\mmts$,  except for a brief era 
near the beginning of the  growth era that can be ignored.\footnote
{This assumption holds within the approximation of  Section \ref{sending}.}
Then  $\chi_k(t)$ at fixed $t$ falls exponentially  
in the regime $k_*(t) \lsim 
 k < a\mmt$ and significant modes have $k\lsim k_*(t)$.

The number of $e$-folds of growth is $N(t)  \equiv
H(t-t\start )$. We denote the end of the linear era 
 by a subscript `end'.
If $N(t\send) \lsim 1$, $k_*(t)$ falls continuously. If instead
$N(t\send) \gg 1$, the exponential increase of $a$ causes $k_*(t)$
to level off after $N(t)\sim 1$.
Using $H\ll \mmt < m$,  we learn that in any case
\be
1  \ll \( \frac{k_*(t)}{a(t\start )H} \)^2 < \frac m H
. \dlabel{kstar1}  \ee
This tells us that the scale
 $k_*(t)$ is  shorter than the  scale leaving the horizon
at the beginning of the waterfall. Since we assume that cosmological
scales are outside the horizon at this stage, $k_*(t)$
is shorter than any cosmological scale. Dividing both sides by
 $\exp(2N(t\send))$  we have 
\be 
 e^{-N(t\send)} \ll  \( \frac {k_*(t)}{a(t\send)H} \) 
\lsim \( \frac m H \)\half
e^{-N(t\send)} \dlabel{kstar}
.\ee
 This tells us that scale $k_*(t)$ (and  in particular, its final value
$k_*(t\send)$)  can be
far outside the horizon at the end of the linear era.

\subsection{Classical field $\chi(\bfx,t)$}

\dlabel{sclassical}

During the growth era 
the mode function $\chi_k$ has constant phase (zero with our convention),
which means that  $\chi_\bfk(t)\propto \chi_k(t)$ can be be regarded as a 
classical field.
The significant modes have $k\lsim k_*(t) \ll  a(t)\mmt$. 
Because  when   for each mode.
 This   means that
the continuous creation of new classical modes, occuring for each mode when
$\omega_k^2$ becomes negative at $k\sim  a(t) m(t)$,
can be ignored.

For the significant modes, $\dot\chi_k/\chi_k \simeq a(t)\mmt$. 
The classical field has approximately the same behaviour;\footnote
{This behaviour breaks down near any locations with $\chi(\bfx,t)=0$.
To discuss them we would
have to extend the discussion to a multi-component $\chi$
as mentioned at the end of Section \ref{sshybrid}. We  assume that if they
exist, they are rare enough to be ignored.}
\be
\dot \chi(\bfx,t) \simeq |m(t)| \chi(\bfx,t) . \dlabel{chidot}
 \ee
Since $k_*(t) \ll a \mmt$, 
 the gradient of $\chi$ is small compared with its time derivative.
 
The spectrum of $\chi$ is   
\be
\calp_\chi(k,t) \equiv(k^3/2\pi^2)P_\chi(k,t)=(k^3/2\pi^2) |\chi_k(t)|^2
\dlabel{chispec} . \ee
 Using \eq{chik} the 
mean square (spatial average) of $\chi^2$ is \cite{p10}
\be 
\vev{\chi^2(t)} =  \int \frac{dk}k \calp_\chi(k,t) =
 \frac1{(2\pi)^3} \int d^3k P_\chi(k,t)  =(2\pi)^{-3/2} P_\chi(0,t)  k_*^3(t)
, \dlabel{classvev} \ee
where $P_\chi(0,t)=|\chi_{k=0}(t)|^2$ is given by \eq{chik1}.

We denote the perturbation in $\chi^2$ by $\delta\chi^2$:
\be
\delta\chi^2(\bfx,t) \equiv \chi^2(\bfx,t) - \vev{\chi^2(t)}
. \dlabel{delchisdef} \ee
The  convolution theorem gives \cite{myaxion}  for 
 $P_{\delta\chi^2}\equiv (2\pi^2/k^3)\calp_{\delta\chi^2}$ 
\be
P_{\delta\chi^2}(k,t)
=\frac{2}{(2\pi)^3}  \int d^3 k' P_\chi(k',t) P_\chi(|\bfk-\bfk'|,t)
. \dlabel{calpchisq}
\ee
For 
 $k\ll k_*(t)$ this gives \cite{p10}
\be \calp_{\delta\chi^2}(k,t) = \frac1{\sqrt\pi} \vev{\chi^2(t)}^2 [k/k_*(t)]^3
. \dlabel{pchisq} \ee
For $k\gg k_*(t)$ it gives
\be \calp_{\delta\chi^2}(k,t) = 2\vev{\chi^2(t)} \calp_\chi(k,t)
, \dlabel{pchisq1} \ee
which falls exponentially at fixed $t$.

\subsection{End of the linear era}

At each location, the linear equation  \eqreff{chidot} 
 ceases to be valid around the time 
when $\chi^2(\bfx,t)$ achieves some value $\chi\snl^2$. 
This time is  given by
\be
\chi^2(\bfx,t\snl(\bfx)) = \chi\snl^2
. \dlabel{tsvevofx} \ee

We will take the linear epoch to end at a time $t\send$,
such that the fraction $y_>$ of space  with $\chi^2(\bfx,t) > \chi\snl^2$
is small.   We will see that 
the probability distribution of $\chi(\bfx,t)$ is gaussian, and
using the approximation $\mbox{erfc}(x)\sim \exp(-x^2)$ we have
\be
\chi\snl^2 \simeq \ln(1/y_>) \vev{\chi^2(t\send)}
. \dlabel{chisvevdef} \ee
According to this equation $t\send$ is not very sensitive to the choice of 
$y_>$,
and for estimates we will take  $\ln(1/y_>)\sim 1$.

If the linear era lasts for long enough, we will have
 $m^2(t\send)\simeq -m^2$ (ie.\ $\phi(t\send)\ll \phi\sub c=m/g$).
 In that case the right hand side
of \eq{fullphi} is irrelevant,  and the linear era ends only when the right
hand side of \eq{fullchi} becomes significant. This gives
\be
\chi\snl^2 \simeq \chivev^2 = 12\mpl^2 H^2/m^2
. \dlabel{chisvev2} \ee

Now suppose instead that $\modm{t\send}\ll m$ (ie.\ $\phi(t\send)\simeq \phi\sub c$).
 Then the linear era will  end
when the right hand side of \eq{fullphi} becomes significant, 
provided that the right hand side
of \eq{fullchi} is then still negligible 
which we are about  to show will be the
case. 
If $\phi$ is still slowly rolling
so that  $3H\dot\phi(t)=-V'$,  
the right hand side of \eq{fullphi} becomes significant. 
when $\chi^2\sim 3H|\dot\phi(0)|/gm $.
But it may be  that $g^2\chi^2$ has become of order $H^2$ first, causing
$\phi$ to  oscillate about the origin.
Including both possibilities we have\footnote
{With  the potential $V(\phi)=\frac12 m_\phi^2\phi^2$ the right hand side
is $\min\{m_\phi^2,H^2\}=m_\phi^2$ which means that only the first possibility
exists. The existence of the second possibility for a more general potential
 was missed in \cite{p10}.}
\be
\chi\snl^2  \simeq \min \left\{\frac{3H\dot\phi(0)}{gm} , \frac{H^2}{g^2} \right\}
. \dlabel{chisvev1} \ee
It  follows from \eqs{phibound}{dotphibound} that 
the right hand side of \eq{chisvev1} is much less than $\chivev^2$,
making the right hand side  of \eq{fullchi} insignificant as advertised.

\subsection{Energy density and pressure of $\chi$}

We have seen that the gradient of $\chi$ is negligible compared with its
time-derivative, and in our adopted gauge the gradient of $\phi$ vanishes.
Ignoring the gradient,
 the energy density and pressure are
$\rho=\rho_\phi+\rho_\chi$ and $p=p_\phi+p_\chi$ where
\bea
\rho_\phi(t) &=&  [V_0 + V(\phi)] +\frac12\dot\phi^2, \dlabel{rphi} \\
p_\phi(t) &=&  -[V_0 + V(\phi)] +\frac12\dot\phi^2, \dlabel{pphi} \\
 \rho_\chi&\simeq& -\frac12 \mmts \chi^2 +\frac12 \dot \chi^2 \dlabel{rchi2a} \\
&\simeq& 0 \dlabel{rchi2} \\
 p_\chi &\simeq&  \frac12 \mmts \chi^2 +\frac12 \dot \chi^2 \dlabel{pchi2a} \\
&\simeq&  \mmts \chi^2  \simeq \dot\chi^2
. \dlabel{pchi2} \eea

In an unperturbed universe the energy continuity equation holds;
\be
\dot\rho(t) = - 3H(t) \( \rho(t) + p(t) \)
. \dlabel{econ} \ee
To the extent that  spatial gradients are  negligible it holds at each location.
With a generic choice of the slicing, denoted by a subscript $g$,  we have
\be
\dot\rho\sub g(\bfx,t) \simeq  - 3\frac{d a\sub g(\bfx,t)}{dt} 
 \( \rho\sub g(\bfx,t) + p\sub g(\bfx,t) \)
,\ee
where $a\sub g(\bfx,t)$ is the locally defined scale factor. We are working in 
the gauge defined in Section \ref{swater}, which means that $t$ is proper time
and we can choose $a(\bfx,t)=a(t)$, the unperturbed scale factor. We therefore
have
\be
\dot\rho(\bfx,t) \simeq  - 3H(t) \( \rho(\bfx,t) + p(\bfx,t) \)
. \dlabel{econlocal} \ee

We are dealing with the linear era, which means that the right hand sides
of \eqs{fullphi}{fullchi} are negligible. With its right hand side
negligible, \eq{fullphi} describes a free field which means that it 
 satisfies the energy continuity equation by itself; 
\be
\dot\rho_\phi \simeq  - 3H\(\rho_\phi + p_\phi\) = - 3H\dot\phi^2
. \dlabel{dotrphi}\ee
The same must therefore be true for $\chi$;
\be
\dot\rho_\chi(\bfx,t) \simeq  - 3H(\rho_\chi(\bfx,t)+ p_\chi(\bfx,t)) 
=-3H\dot\chi^2(\bfx,t)
. \dlabel{econ2} \ee

At each location we have\footnote
{In the present context \eq{chiddot2} can be replaced by \eq{chidot}
 by virtue of the adiabaticity condition. But 
in Section \ref{smconst} we drop the adiabaticity condition.}
\be
\ddot \chi(\bfx,t) + 3H\dot\chi(\bfx,t) + m^2(t) \chi(\bfx,t) \simeq 0
. \dlabel{chiddot2} \ee
Using this equation  to differentiate \eq{rchi2a} we find
\be
\dot\rho_\chi(\bfx,t) \simeq  - 3H(\rho_\chi(\bfx,t)+ p_\chi(\bfx,t)) -\frac12
\frac{d\mmts}{dt} \chi^2(\bfx,t)
. \dlabel{dotrhochi}  \ee
The second term of the right hand side 
violates the energy continuity equation. This apparent inconsistency
between the field equations and the energy continuity equation  occurs because
the effect of the interaction term $g^2\phi^2\chi^2$ is dropped in
\eq{fullphi} (ie.\ the right hand side is set to zero) but kept in \eq{fullchi}.

If we demand approximate
consistency between the field equations and the energy continuity equation,
we need the second term of \eq{dotrhochi} to be much smaller than the first.
That condition is equivalent to
\be
\frac{d\mmt}{dt}  \ll \mmt H
, \dlabel{condition} \ee
which  is stronger than the adiabaticity conditions  \eqsref{adcon2}{adcon3}
(with $k\ll a\mmt$).
But there is no need to impose this
stronger condition,  because the near cancellation of the two terms of $\rho_\chi$
makes it unreasonable to expect the approximate evolution \eqreff{chiddot2} of $\chi$
to give even an 
approximate estimate of $\rho_\chi$. By contrast, the right hand side of
the energy continuity equation has no cancellation so that it can be used to evaluate
$\rho_\chi$. Invoking \eq{chidot}, we find
\be
\rho_\chi(\bfx,t) \simeq  
-\frac32 H \mmt \chi^2(\bfx,t) \dlabel{rchi3}
. \ee

\subsection{Justifying the neglect of $\overline \chi$}
\dlabel{sjust}

It has been  essential for our discussion that the spatial average of $\chi(\bfx,t)$
is negligible. That is the case in a sufficiently large volume, because $\chi(\bfx,t)$
is constructed entirely from the Fourier modes. 
But to  make contact with cosmological observations we  should  consider a finite box,
whose  (coordinate) size  $L$ is not too many orders of magnitude bigger than
 the size of the presently observable universe
 \cite{mybox}.
Denote the average within the box  by  $\bar\chi$ we have
\be
\chi(\bfx) \simeq  \bar\chi + \chi_>(\bfx)
, \ee
where the Fourier modes of $\chi_>$ satisfy $kL>1$ so that
\be
\vev{\chi_>^2} = \int^\infty_{L\mone} \frac{dk}k \calp_\chi(k) 
 \dlabel{i1}
. \ee

The average within the box comes from modes with $kL\lsim 1$, 
and for a random location 
of the box  the expectation value of $\bar\chi^2$ is
\be
\vev{\bar\chi^2}  \simeq  \int^{L\mone}_0 \frac{dk}k \calp_\chi(k)  \dlabel{i2} 
. \ee

To justify the neglect of $\bar\chi$ we need
$ \vev{\bar\chi^2}\ll \vev{\chi_>^2} $.
In our scenario, where $\calp_\chi(k)$ peaks at a value $k_*(t)$, this 
is equivalent to  $Lk_*(t) \gg 1$. That is satisfied because the scale
  $k_*(t)$ is supposed to be much smaller than the observable universe.
To have $\vev{\bar\chi^2} \gsim  \vev{\chi_>^2}$  we would
 presumably have to allow the transition from $m^2(t) = -H^2$
to $m^2(t) = H^2$  to take at least several $e$-folds so that it 
can generate a contribution to $\chi$ that has a flat spectrum.
Then, if the flat spectrum generated during the transition dominates,
one would have
\be
\vev{\bar\chi^2}/\vev{\chi_>^2} \simeq N\sub{before}/N\sub{after}
, \ee
where $N\sub{before}$ 
($N\sub{after}$) is the number of $e$-folds of transition before 
(after) the
observable universe leaves the horizon. 

\section{Contribution $\zeta_\chi$ to the curvature perturbation}

\dlabel{scurv}

We write the  contribution to the curvature perturbation that is generated during
the waterfall as $\zeta_\chi = \zeta\slin+\zeta\snl$, where the first term is
generated during the linear era, and the second is generated afterward up to
some  epoch just after inflation has ended. 

The curvature perturbation is 
$\zeta(\bfx,t)\equiv \delta \ln a(\bfx,t)$, where $a(\bfx,t)$ is the locally
defined scale factor on the spacetime slicing of uniform $\rho$.
As in \eq{econlocal}, the spatial gradient is supposed to be negligible,
which in general requires smoothing on a super-horizon scale. 
Using that equation, we see that the 
change in $\zeta$ between any two times $t_1$ and $t_2$
is 
\be
\zeta(\bfx,t_2) - \zeta(\bfx, t_1) =\delta N(\bfx,t_1,t_2)
, \ee 
where $N(\bfx,t_1,t_2)$ where $N$ is the 
number of $e$-folds between slices of  uniform $\rho$. 
Working to first order in $\delta\rho$,\footnote
{A  second-order calculation of  $\zeta$ is   needed only to treat very small
non-gaussianity corresponding to reduced bispectrum $|\fnl|\lsim 1$.
 On cosmological scales, such non-gaussianity
 will eventually be measurable (and is expected  if $\zeta$
comes from a curvaton-type  mechanism \cite{book}).
  But there is no hope of detecting such   
non-gaussianity on  much smaller scales.}
we will use this result  to calculate $\zeta\slin$, and then see how it
might be used to calculate $\zeta\snl$.

We note in passing that an equivalent procedure is to
integrate the expression
\be
\dot\zeta(\bfx,t) =  -\frac{H(t)}{\rho(t) + p(t)} \delta p\sub{nad}(\bfx,t)
\dlabel{zetadot}, \ee
where 
\be
 \delta p\sub{nad}(\bfx,t) \equiv  
\delta p(\bfx,t) - \frac{\dot p(t)}{\dot\rho(t)} \delta\rho(\\
bfx,t)
. \dlabel{deltapnad} \ee

\subsection{The contribution  $\zeta\slin$}

\dlabel{sszetalin}

During the linear era, the gradient of $\rho$ is negligible without any 
smoothing.
We   are working
in a gauge where $\delta\phi=0$ 
so that $\rho(\bfx,t)=\rho_\chi(\bfx,t) + \rho_\phi(t)$.
Ignoring the inhomogeneity of the locally defined Hubble parameter,
\be
\zeta\slin(\bfx,t) =  H \[ \delta t(\bfx,t\send) - \delta t(\bfx,t\start ) \]
, \dlabel{zchi} \ee
and 
\be
H\delta t(\bfx,t) \equiv - H\frac{\delta\rho_\chi(\bfx,t)}{\dot\rho(t)}
=\frac13 \frac {\delta\rho_\chi(\bfx,t)}{\vev{\dot\chi^2(t)} + \dot\phi^2(t) }
. \dlabel{deltat} \ee

Using \eq{rchi3},
\bea
H\delta t(\bfx,t) &=& 
\frac13 \frac 
{
\rho_\chi(t)
}{
\vev{\dot\chi^2(t)} 
}
\frac{
\vev{\dot\chi^2(t)}
}{
\vev{\dot\chi^2(t)} + \dot\phi^2(t) 
}
\frac{
\delta\chi^2(\bfx,t)
}{
\vev{\chi^2(t)}
} \dlabel{deltat2} \\
 &=& 
-\frac12 \frac {H}{\mmt} 
\frac{
\vev{\dot\chi^2(t) }
}{
\vev{\dot\chi^2(t)} + \dot\phi^2(t) 
}
\frac{
\delta\chi^2(\bfx,t)
}{
\vev{\chi^2(t)}
} \dlabel{deltat3} 
. \eea
Using \eq{pchisq}, we have for $k\ll k_*(t)$
\be
H^2\sqrt\pi \calp_{\delta t}(k,t) =  \( \frac H{2\mmt} \)^2
\( \frac{
\vev{\dot\chi^2(t) }
}{
\vev{\dot\chi^2(t)} + \dot\phi^2(t)
} \)^2
\( \frac k{k_*(t)} \)^3 
. \dlabel{pzetachi9} \ee

We  assume that $|\delta t(t\send)| \gg |\delta t(t\start )|$, which will be
justified within the approximation of Section \ref{sending}.
 Then  we have
\bea
\zeta\slin(\bfx) &=& - H \delta t(\bfx,t\send) \dlabel{zetachifirst} \\
&=&  -\frac {H}{2\modm{t\send} }
\frac{
\vev{\dot\chi^2(t\send)}
}{
\vev{\dot\chi^2(t\send)} + \dot\phi^2(t\send) } 
 \frac{
\delta \chi^2(\bfx,t\send)
}{
\vev{\chi^2(t\send)} 
} \dlabel{zetachi}
. \eea 
The inhomogeneity of $H$ is indeed negligible because 
 it generates a contribution
\be  \int^{t\send}_{t\start } \delta H(t,\bfx) dt = \frac H{2\rho_\chi}
\int^{t\send}_{t\start } \delta\rho(\bfx,t)  dt
\simeq \frac{H \delta\rho_\chi(\bfx,t\send)}{2\rho \modm{t\send}}
.  \ee
This is much less than $H\delta t(t\send)$ in magnitude, because $\mmt \gg H$ and
 $|\dot\rho|\ll H\rho$.

Using \eq{pzetachi9}, we have for $k\ll k_*(t\send)$
\be
\sqrt\pi \calp_{\zeta\slin}(k) =  \( \frac H{2\modm{t\send} } \)^2
\( \frac{
\vev{\dot\chi^2(t\send) }
}{
\vev{\dot\chi^2(t\send)} + \dot\phi^2(t\send)
} \)^2
\( \frac k{k_*(t\send)} \)^3  \ll \( \frac k{k_*(t\send)} \)^3
. \dlabel{pzetachi} \ee
In the opposite 
regime  
$k\gg k_*(t\send)$, $\zeta\slin$ is  negligible  because $\delta\chi^2$
is.  Therefore, $\calpzlin$ peaks at $k_*(t\send)$ with the value
\be
\sqrt\pi \calp_{\zeta\slin}(k_*(t\send)) \simeq   \( \frac H{2\modm{t\send} } \)^2
\( \frac{
\vev{\dot\chi^2(t\send) }
}{
\vev{\dot\chi^2(t\send)} + \dot\phi^2(t\send)
} \)^2
. \dlabel{pzetachipeak} \ee

If $m^2(t\send)\simeq -m^2$, \eq{chisvev2} holds and we have
\be
\vev{ \dot\chi^2(t\send) }/\dot\phi^2(t\send) 
\simeq  12\mpl^2H^2/\dot\phi^2(t\send) \gg 1
.\dlabel{chidotphidot} \ee
(The inequality follows from  \eq{dotphibound}). Then 
 \eq{pzetachi} simplifies to
\be
\sqrt\pi \calp_{\zeta\slin}(k) =  \( \frac H {2m} \)^2
\( \frac k{k_*(t\send)} \)^3
. \dlabel{pzetachi1} \ee

\subsection{The contribution   $\zeta\snl$}

\dlabel{sszetasvev}

%%\begin{figure}[htb]
%%\begin{figure}
%\FIGURE{
%\centering
%\includegraphics[width=0.6\columnwidth]{figure1.eps}
%\caption{The `end-of-inflation' formula
% gives the change in $\zeta$, generated by a sufficiently fast transition
%between nearly-exponentila inflation and non-inflation. For the formula
%to apply, we need $\Delta t(\bfx)\ll |\delta t_{12}|$ at a typical position.
%Figure courtesy of T.\ Matsuda.}
%\label{one}
%}
%%\end{figure}

 Let us estimate the number of $e$-folds 
$N\snl$ after the end of the linear era.
At $t\send$, $\chi^2$ is increasing exponentially. Soon afterward it starts
to affect  $\phi$, driving it towards zero. We therefore expect 
 $m^2(t)$ to quickly approach $-m^2$ 
after $t\send$ (if it is not there already),
 restoring at least approximately
the linear evolution of $\chi^2$. Then \eq{chidot} will hold with
$\mmt\sim m$   giving
\be
N\snl \sim (H/m)  \ln(\chivev/\chi\send)
. \dlabel{deltaN} \ee
If the linear era ends only when the right hand side of \eq{fullchi}
becomes important we have $\ln(\chivev/\chi\send)\sim 1$,
giving $N\snl \sim H/m \ll 1$. But 
if it ends when the right hand side of \eq{fullphi} becomes important 
we may have  $\ln(\chivev/\chi\send)\gg  1$ which allows $N\snl \gsim 1$.

Now we consider the contribution $\zeta\snl$, that is generated
between  $t\send$ and some time $t_2$ just after inflation has
everywhere ended.  To calculate it we need to smooth on a super-horizon scale.
Then we can use the $\delta N$ formula which gives
\be
\zeta\snl(\bfx,t) = H\delta t_{12}(\bfx)
, \dlabel{zetasnl} \ee
where the initial and final slices both have uniform $\rho$ and 
$t_{12}(\bfx)$ is the
proper time interval between them.

At each location, the linear era ends at the epoch $t\snl(\bfx)$
given by \eq{tsvevofx}. At this epoch  there is nearly-exponential inflation, 
and  inflation ends  at some later time $t\sub{noninf}(\bfx)$.
If $ \Delta t(\bfx) \equiv t\sub{noninf}(\bfx)-t\snl(\bfx)$
is sufficiently small it can be  taken to correspond to  a spacetime slice
of negligible thickness. Then $\delta t_{12}$ is given by the
`end-of-inflation' formula   \cite{endinf} 
\be
\delta t_{12}(\bfx)
\simeq  \frac{\delta \rho(\bfx)}{\dot\rho(t\send)}
, \dlabel{einf} \ee
where $\delta \rho(\bfx)$ is defined on the slice.
The addition of $\zeta\snl$ to $\zeta\slin$ corresponds to taking 
the final slice of the $\delta N$ formula be the transition slice, instead of
a slice of uniform $\rho$. 

This equation is valid to first order in $\delta\rho$. 
To derive  it  we take the  separation between the 
initial and final slice to be not much bigger than is needed for them to
enclose the transition slice.
Then we  can take the unperturbed quantity
$\dot\rho(t)$ to have a 
constant value both during inflation and non-inflation. This gives
\be
\delta t_{12}(\bfx) \simeq  \delta\rho(\bfx)\(\frac1 {\dot\rho\sub{inf} } 
-\frac 1{\dot\rho\sub{noninf} }\)
. \ee
Since $|\dot\rho\sub{inf}|$ is evaluated during nearly-exponential inflation,
it is much smaller 
than $|\dot\rho\sub{noninf}|$ leading to \eq{einf}.

We are defining $\delta\rho_\chi$ on a slice of uniform $\rho_\phi$
and $\delta\rho$ is defined on a slice of uniform $\chi$.
The time displacement from the first slice to the second slice is
is $ -\delta\rho_\chi/\dot\rho_\chi$, 
which means that 
 $\delta\rho=\delta\rho_\chi \dot\rho_\phi(t\send)/\dot\rho_\chi(t\send)$.
Putting this into \eq{einf} we get
\be
\zeta\snl(\bfx)/\zeta\slin(\bfx) \simeq  
 \dot\rho_\phi(t\send)/\dot\rho_\chi(t\send) \simeq
\dot\phi^2(t\send)/\dot\chi^2(t\send)
. \ee
We are only interested in the case that this ratio is $\gsim 1$.
Then the inclusion of $\zeta\snl$ corresponds to omitting
the middle term of  \eq{zetachi}. {}From \eq{chidotphidot}, this 
case can occur only if $\modm{t\send} \ll m$.

Now comes a crucial point. From the derivation of 
 \eq{einf}, it is clear that the criterion for its validity is 
 $\Delta t(\bfx)\ll |\delta t_{12}(\bfx)|$,  at a typical location. 
(In words, the thickness  of the transition slice is  negligible compared 
compared with its warping.)
This simple remark has not been made before, and
consequently it has not been checked 
 whether  the criterion is satisfied. 

In our case, $H\Delta t(\bfx)$ is given by \eq{deltaN}
with $\chi\send^2$ replaced by $\chi\snl^2(\bfx)$. Since that quantity
appears only in the log the change will not have much effect,
and we will have $H\Delta t(\bfx)\sim N\snl$ at a typical location.
On the other hand, the typical value of $|\zeta\snl(\bfx)|
=  H |\delta t_{12}(\bfx)|$ is $\calp\half_{\zeta\snl}(k)$
where $k\lsim aH$ is the smoothing scale used to define
$\zeta\snl$. The criterion for \eq{einf} to be valid
is therefore $N\snl \ll \calp\half_{\zeta\snl}(k)$. In the regime
of interest $\dot\phi^2(t\send)/\dot\chi^2(t\send) \gg 1$, this criterion
becomes
\be
\frac{\modm{t\send}} m\ln \(\frac {\chivev}{\chi\send} \) \ll \(
\frac k{k_*(t\send)} \)^{3/2}
.  \dlabel{criterion}. \ee
Whenever the criterion \eqreff{criterion} is not satisfied,
the calculation of $\zeta\sub{end}$ that we have described does not apply.
%An alternative approach might be to suppose that the approximation
%$m^2(t)\simeq -m^2$ becomes valid very soon after $t\send$.
%Then the  linear evolution of $\chi$ would be quickly re-established
%and it might be possible to extend the calculation of $\zeta\slin$
%beyong $t\send$ arriving at something like case 2. We have not investigated
%that possibility. 

\subsection{Other uses of the `end of inflation' formula}

\dlabel{sother}

Our use of \eq{einf} to evaluate $\zeta\snl$
is quite different from its usual applications
\cite{endinf,endapps1,endapps2,endapps3,endapps4}.
In those applications, 
the field causing $\delta\rho(\bfx)$
has  a nearly flat spectrum, leading to a nearly flat
 $\calp_{\zeta\snl}(k)$ that can  give  a significant
(even dominant) contribution to $\calpz(k)$ on cosmological scales.
Since $\calpz\half(k) \sim 5\times 10\mfive$ on these scales,
 \eq{criterion} on cosmological scales  becomes
\be
N\sub{tran} \ll  \calp_{\zeta_{12}}\half < 5\times 10\mfive
, \dlabel{criterion2}. \ee
where $N\sub{tran}$  now refers to the duration of the transition slice
in the scenario under consideration, and $\zeta_{12}=H\delta t_{12}$ is the
contribution to $\zeta$.

Most of the other  applications \cite{endinf,endapps1,endapps2}
consider hybrid inflation, with 
 the transition slice  the entire
hybrid inflation waterfall. Of course their setup is different from ours
because they introduce a third field, the one that generates $\delta \rho$
in \eq{einf}. 
In these cases, $N\sub{tran}$ 
in \eq{criterion2} becomes  the total
duration of the waterfall. We see from \eq{deltaN} that it cannot
be much less than $H/m$, which means that \eq{criterion2}
needs $H/m\ll 5\times 10\mfive$. Since we need $(H/m)^2\gg H/\mpl$
(corresponding to $\lambda\ll 1$), this requires a fairly low inflation
scale $H\ll 10^{-9}\mpl$.

An alternative possibility  \cite{endapps3} 
is for the transition slice to be at the end
of thermal inflation \cite{thermal,therm2,therm3,therm4}. 
(Thermal inflation is is a few $e$-folds of inflation occurring typically
long after the usual inflation, which is ended by a thermal  phase transition.)
 Then we expect roughly
$\Delta N\sim H/m$, where
 $m$ is the tachyonic mass of the field  causing the end of
thermal inflation. This  criterion \eqreff{criterion2} is satisfied by the usual realizations
of thermal inflation. 
Further possibilities for the transition slice are considered
in \cite{endapps4}.

\section{Effect of $\zeta_\chi$}

\subsection{Cosmological black hole bound on $\calpz$}

\dlabel{sbh}

The most dramatic effect of $\zeta$ would be the formation of
black holes. This places an upper bound on $\calpz$, which we now
discuss taking on board for the first time the non-gaussianity of $\zeta$.

The bound that we are going to consider 
rests on the validity of the following statement: if,
 at any epoch after inflation,
there are  roughly spherical and horizon-sized regions with $\zeta$
significantly bigger than 1, a significant fraction of them  will collapse
to form roughly horizon-sized   black holes.\footnote
{We are choosing the background scale factor $a(t)$ so that
the perturbation $\zeta=\delta(\ln a(\bfx,t))$ has zero spatial average.}
The validity is suggested by 
 the following argument: the overdensity at horizon entry
is $\delta\rho/\rho \sim \zeta$, and if it is of order 1 then
 $\delta \rho \sim \rho=3\mpl^2H^2$. The excess energy within the Hubble distance
 $H\mone$
is then  $M\sim H\mthree \rho \sim \mpl^2/H$, which means
that the 
Hubble distance corresponds roughly to the Schwarzchild radius of a black
hole with mass $M$. The validity 
is confirmed by detailed calculation using several
different approaches, as summarized for instance in \cite{bhbound}.

Before continuing we mention the following caveat. Practically all of the literature,
as well as the simple argument just given, assumes
 that $\zeta$ within the region is not {\em very much} bigger than
1. Then the spatial geometry within the region is not too strongly distorted
and the size of the black hole is indeed roughly that of the horizon. 
In the opposite case, the background geometry is strongly distorted and the wavenumber
$k$ defined in the background no longer specifies the physical size of the region
at the epoch $aH=k$ of horizon entry   \cite{bhform2}. 
An entirely different discussion would then
be necessary, which has not been given in the literature. As the opposite case does
not arise in typical early-universe scenarios we ignore it.

We are interested in the case that $\calpz(k)$ has a peak at some value
$\kpeak$, and we assume that the width of the peak in $\ln k$ is roughly of 
order 1 so that 
\be
\vev{\zeta^2} =  \int^\infty_0 \calpz(k) dk/k \simeq \calpz(\kpeak)
. \dlabel{kcubed} \ee
Regions with $\zeta\gsim 1$ that might
form black holes will be rare if  $\calpz(\kpeak)$ is not too big.
 Observation demands that the regions must indeed be rare, because
it places a strong upper bound 
on the fraction of  of space that can collapse 
to form horizon-sized 
black holes, on the assumption that the collapse takes place at a single
epoch as is the case in our scenario. 
A recent investigation of the bound  is given in 
\cite{bhbound}, with extensive references to the literature. The bound 
 depends  on the epoch of collapse. Denoting it by $\beta$
it lies in the range
\be
10^{-20} \lsim \beta \lsim 10\mfive
. \dlabel{betabound} \ee
To bound $\calpz(\kpeak)$, we 
 shall require $y<\beta$ where $y$ is the  fraction of space 
with  $\zeta>\zeta\sub c$, and $\zeta\sub c$ is roughly of order 1.

The fraction $y$ can be calculated from $\vev{\zeta^2}$
if we know the probability distribution of $\zeta(\bfx)$. The standard 
assumption is that it is gaussian. Then
\be
y = \frac12 \mbox{erfc}\, (\zeta\sub c/\sqrt {2 \vev{\zeta^2}} )
, \ee
and using the large-$x$ approximation $\mbox{erfc}\,(x)\simeq e^{-x^2}
/\sqrt\pi x \sim  e^{-x^2}$
we find
\be
\calpz(k\sub{peak}) \simeq \vev{\zeta^2} \lsim \zeta\sub c^2/2 \ln(1/f)
. \ee  
For the range \eqreff{betabound} this gives (with  $\zeta\sub
c\simeq 1$)  $\calpz(\kpeak)\lsim 0.01$  to $0.04$.

But $\zeta\slin$ given by \eqs{delchisdef}{zetachi} is actually non-gaussian, of the 
form 
\be
\zeta = -(g^2 - \vev{g^2} ) \dlabel{zetang}
. \ee
With this form, there 
is    no region of space where $\zeta > \vev{g^2}$, and 
 $y\ll 1$ now implies some bound $\vev{g^2}-\zeta\sub c\ll \zeta\sub
c$
which is practically equivalent to  $\vev{g^2}< \zeta\sub c$.
This  corresponds to $\calpz(\kpeak)\lsim  2$
\be
\calpz(k\sub{peak}) \simeq  \vev{\zeta^2} = 
2\vev{g^2}^2\lsim 2 \zeta\sub c^2 \lsim 2
. \ee

For completeness, we see what happens if
$\zeta = +\( g^2 - \vev{g^2}\)$ with $g$ gaussian. 
(This might be the case
\cite{ourbh}  if $\zeta$ is  generated after inflation by a curvaton-type
 mechanism.)
The we have 
\be
\calpz(k\sub{peak}) \sim
\vev{\zeta^2} =  2\vev{g^2}^2 \lsim  2 \[ \frac{\zeta\sub c}{2 \ln(1/y) } \]^2
, \ee
which  gives $\calpz(\kpeak)\lsim 2\times 10\mfour$  to 
$2\times 10\mthree$.

In all three cases, the bound on $\calpz(k\sub{peak})$ is very insensitive
to $f$ which means that it depends only weakly on the value of $\beta$.
Turning that  around though, the black hole abundance is very sensitive to
$\calpz(k\sub{peak})$ which  suggests that fine-tuning of parameters will
be needed to get an eventually observable (yet presently allowed) abundance.

If the peak has width $\Delta \ln k$ different from 1, $\vev{\zeta^2}
\simeq \calpz(k\sub{peak})\Delta \ln k$. If $\Delta \ln k\ll 1$
this  weakens the bound on $\calpz(k\sub{peak})$ by a factor
$(\Delta \ln k)\mone$, but such a narrow peak is not generated in typical
scenarios. If instead $\Delta \ln k\gg 1$, one might think that the bound
on  $\calpz(k\sub{peak})$ is strengthened by a factor
$(\Delta \ln k)\mone$, but that conclusion is too hasty because the 
observational bound \eqreff{betabound} refers to the formation of horizon
sized black holes  at a more or less 
definite epoch whereas the broad peak will 
lead to the formation of such black holes
over $\Delta \ln k$ Hubble times. The value of $\vev{\zeta^2}$ in that case
is not directly related to the black hole abundance, and the black hole bound
on $\calpz(k\sub{peak})$  is unlikely to be
strengthened very much.
For instance, if the observational bound on black hole abundance applies
 separately to the black holes formed within each unit interval of $\ln k$,
 the effective value of $y$ for a given value of
$\calpz(k\sub{peak})$ is  just multiplied by that 
factor,  which has  a negligible effect on the bound on
 $\calpz(k\sub{peak})$. 

\subsection{The effect of $\zeta\slin$}

\dlabel{seffect}

Now we discuss the effect of $\zeta\slin$, assuming that it is at least
not canceled by $\zeta\snl$.
 By virtue of \eq{adcon2}, the first term of \eq{pzetachi} is $\ll 1$,
and the second term is $\leq 1$.
If   $k_*(t\send)$ is super-horizon, 
 $\zeta$ is of the form \eq{zetang} with the minus sign, and the
 black hole bound is $\calpz\lsim (k_*(t\send))\lsim  2$. This is likely
to be well satisfied.

If instead $k_*(t\send)$ is sub-horizon, 
we have to remember that the black hole bound
refers to  horizon-sized regions. To apply it, we must 
 drop sub-horizon  modes of $\zeta\slin$. Estimating the bispectrum, trispectrum 
as in \cite{p10}, one sees that this makes $\zeta\slin$ nearly gaussian.
Then $\calp_{\zeta\slin}$ peaks at $k\sim k\sub{end}\equiv a(t\send) H$, and the 
 black hole bound is   roughly
$\calpzlin(k\sub{end})\lsim 10\mtwo$. This too will be  satisfied if
 $k_*(t\send)$ is well within the horizon.

We  emphasize that these  
bounds refers to the formation of {\em horizon-sized} 
black holes. 
If $k_*(t\send)$ is sub-horizon, smaller black holes may also be formed. 
A  discussion of their abundance would require assumptions about the evolution
of the perturbations during the transition from inflation to non-inflation,
and would be much more
difficult than the corresponding discussion
\cite{subhor} for the formation of black holes from $\zeta_\phi$.

Although $\calpzlin(k)$ is probably
too small to form black holes, it may still be
quite large. If reheating after inflation is long delayed this may lead
to copious structure formation with a variety of possible cosmological effects
\cite{efg}.

Finally, let us see whether $\calpzlin(k)$ 
can be significant on cosmological scales;
ie. whether it can be comparable with the observed quantity
$\calpz\simeq 10^{-9}$. It follows from \eq{kstar} that the 
scale $k_*(t\send)$ is shorter than the scale leaving the horizon 
at the beginning of the waterfall. Therefore, the inequality \eqreff{pzetachi}
implies that $\calpzlin$ will give a negligible 
 contribution to the observed quantity $\calpz\sim 10^{-9}$
if the shortest cosmological scale leaves the horizon more than $3\ln(10) 
\simeq 7 $ $e$-folds before the start of the waterfall, ie.\ if the 
observable universe leaves the horizon more than $\simeq 22$ $e$-folds before
the start of the waterfall. We will see that this is assured within the
approximation of Section \ref{sending}.

\section{Estimates using a simple approximation}

\dlabel{sending}

In this section we make a simple  approximation for $m^2(t)$.
 This will
 allow us to verify some of the assumptions that we have been making,
especially if we assume that $\phi$ satisfies the slow-roll approximation.

\subsection{The approximation for $m^2(t)$}

The approximation is 
\bea
m^2(t) &\simeq& - \mu^3 t\qquad (0 \lsim \mu^3t  <  m^2) \dlabel{approx1} \\
m^2(t) &\simeq& - m^2 \qquad (\mu^3t >  m^2), \dlabel{approx2}  \\
\mu^3&\equiv&  2gm |\dot\phi(0)|
. \dlabel{msqexp} \eea
 The  cross-over between the two expressions 
is at  $t=t\smax\equiv m^2/\mu^3$. The second expressions
corresponds to  setting $\phi=0$. If the linear era ends at $t< t\smax$
only the first approximation is invoked.

The first approximation is exact at $t=0$, and
it ignores the time-dependence of     $d(\phi^2)/dt =2\phi\dot\phi$.
The constancy of $\phi$ is a good approximation at $t\ll t_=$,
and so is the constancy of $\dot\phi$ if
 \eqreff{ddotphibound}  is sufficiently well satisfied.
The  fast transition requirement described in Section \ref{sstandard} is
$H\lsim \mu$. To simplify some of the estimates we will usually take the
requirement to be 
\be
H\ll   \mu\mbox{\ \ \ \ (fast transition)}
. \dlabel{fasttran} \ee

 With this  approximation for $\mmts$, the linear era is completely described
by  the 
four parameters  $g$, $H$, $m$,  and $\mu$.
 Let us define $N(t)\equiv Ht$. Then the epoch $t=t_=$ corresponds to
\be
 N(t\smax) = \( \frac m\mu \)^2 \frac H\mu = \( \frac m\mu \)^3 \frac Hm
=\( \frac mH \)^2 \( \frac H\mu \)^3
. \dlabel{nmaxdef} \ee

\subsection{Slow-roll approximation}

\dlabel{ssingle}

To obtain the strongest possible results, we assume that 
 the evolution of $\phi$ satisfies the slow-roll 
approximation, at least during some era that begins before the waterfall and ends
when $\phi$ ceases to to affect the evolution of $\chi$. 

Then  unperturbed  inflaton field satisfies
\be
\ddot\phi(t) + 3H(t) \dot\phi(t) + V'(\phi(t)) = 0
\dlabel{phieq} .  \ee
The basic slow-roll approximation is 
\be
3H\dot\phi\simeq  - V'(\phi) 
, \dlabel{dotphi} \ee
or equivalently
\be
H|\ddot\phi/\dot\phi| \ll 1 \dlabel{ddotphibound}  
. \ee
The requirement that  the first derivative  of \eq{dotphi} be 
 consistent with \eq{ddotphibound} is
$\epsilon_H +  \eta\ll 1$
where $\epsilon_H\equiv |\dot H|/H^2$ and
$\eta\equiv  V''/3H^2$. The slow-roll approximation assumes
 $\epsilon_H\ll 1$ and $|\eta|\ll 1$.

Before the waterfall, and for as long afterward as 
$\chi$ has a negligible effect on $\phi$, the 
 Fourier components of the perturbation $\delta\phi$ satisfy
\be
\ddot \delta\phi_\bfk + 3H(t)\delta\phi_\bfk + (k/a)^2
\delta\phi_\bfk + V''(\phi(t))\delta\phi_\bfk  = 0
. \ee
This equation ignores back-reaction, which is a good approximation by
 virtue of the slow-roll approximation  \cite{book}.

As a  scale leaves the horizon, 
 the vacuum fluctuation of $\phi$
is converted to a classical perturbation $\delta\phi$
 with spectrum $\simeq (H/2\pi)^2$.
At a given epoch, the 
 vacuum fluctuation is set to zero on sub-horizon scales.

These results hold both before and during the waterfall. 
Focusing on the former era  we have more results,
because $\phi$ is the only field.  First, we have a couple more
 relations:
\bea
3\mpl^2 H^2(t) &=& \rho_\phi \simeq  V \equiv V_0 + V(\phi) \dlabel{hexp} \\
\epsilon_H &=& \frac12 \frac{\dot\phi^2}{\mpl^2H^2} \simeq
  \epsilon\equiv \mpl^2 (V'/V)^2/2 \dlabel{eexp}
. \eea
Second, we have the crucial result that  $\delta\phi$ generates
 nearly gaussian curvature perturbation
$\zeta_\phi$ to the curvature perturbation $\zeta$, with spectrum given by
\be
\calpzphi\half(k) \simeq  \left. \frac{H^2}{2\pi\dot\phi} \right|_{aH=k}
. \dlabel{calpzphi} \ee
For a given $k$ the spectrum is generated at the epoch of horizon 
exit $k=aH$, and is constant thereafter until at least the beginning of 
the waterfall.

\subsection{Trading $\mu$ for $f$}

\dlabel{strading}

For single-field inflation, we  can use   \eq{calpzphi} to  
 obtain more powerful results by  trading $\mu$ for
\cite{p10}
\be
 f \equiv
 \(5\times 10^{-5}\)\mone H^2/2\pi\dot\phi(0)  = \(5\times 10^{-5}\)\mone
\calp_{\zeta_\phi}\half(k\sub{beg})
, \dlabel{fdef} \ee
where $k\sub{beg}$ is the horizon scale at the beginning of the waterfall.
Inflation models are usually constructed so that $ \calp_{\zeta_\phi}$
accounts for the observed  $\calpz$ on cosmological scales. Then, if
$\calp_{\zeta_\phi}$  is nearly scale-independent we will have
$f\sim 1$. More generally there is an upper bound
\be  f\lsim 2\times 10^3\mbox{\ \ \ \ (black hole constraint)} \dlabel{bhole}
 \ee
corresponding to  the black hole 
bound $\calp_{\zeta_\phi} \lsim 10\mtwo$ on the spectrum of the nearly
gaussian $\zeta=\zeta_\phi$ that exists at the beginning of the 
waterfall.  There is also a lower bound
corresponding to
 \eq{dotphibound}:
\be
f\gg 10\mtwo  H/(10\mfive \mpl)\mbox{\ \ \ \ (nearly exponential inflation)}
. \dlabel{fgg} \ee

The relation between $f$ and $\mu$ is given by
\be
\( \frac H \mu \)^3 \simeq 10\mfour \frac{fH}{gm}
. \dlabel{sudden} . \ee
We are demanding $g\ll 1$, but the 
 fast transition requirement $H\ll \mu$ can always be satisfied because
$f<2\times 10\mthree$ and $H\ll m$. 
%But \eq{fgg} implies a lower bound
%on $H/\mu$;
%\be
%\( \frac H \mu \)^3 \gg  10^{-6} \frac Hm \frac H{10\mfive \mpl} 
%. \dlabel{sudden2} \ee

In this paper we are not  specifying the potential
$V(\phi)$. Most previous work considers the potential $V(\phi)=\frac12
m_\phi^2 \phi^2$. Then slow-roll requires $m_\phi\ll H$
and  $3H\dot\phi=-m_\phi^2 \phi$.   
The fast  transition requirement $H\lsim \mu$ becomes
\be
\( \frac{m_\phi^2}{H^2} \) \( \frac{m^2}{H^2} \) \gsim  1
, \dlabel{fastquad} \ee
and $f$ is given by
\be
f=10^4 g 
\( \frac{m_\phi^2}{H^2} \)\mone \( \frac{m}{H} \)\mone
. \ee
In this case we need $f\ll 1$, to avoid
 a positive spectral tilt for $\calpz$
which would  conflict with observation.
The requirement that  $V(\phi)$ does not support inflation
(so that inflation ends with the end of the waterfall)
is $\phi\sub c\lsim 10\mpl$, which is guaranteed by \eq{phibound}. 

\subsection{The case $t\send < t\smax$}

\dlabel{scase1}

In this case $m^2(t) \sim  -\mu^3t$, and \eq{chisvev1} becomes
\be 
\chi\snl^2\simeq \min \left\{ \frac{3H\mu^3}{2g^2m^2}, \frac{H^2}{g^2} \right\}
. \dlabel{chinl2} \ee
Using \eq{chisvev1} and  $\dot\chi^2  \simeq  |m^2(t)| \chi^2$,
\be
\frac{\vev{ \dot\chi^2(t\send)  } }{\dot\phi^2(t\send) }
\simeq  
\min \left\{ 6N(t\send), 4N(t\send)\frac{g^2H m^2}{\mu^3} \right\} 
. \dlabel{6htend} \ee

Growth begins when
 \eq{adcon3} is satisfied, corresponding to $\mu t_1 \simeq 1$, and
 $\mu t\sub{start} = 2\mu t_1 \simeq 2$ and $Ht\sub{start}\ll 1$.

The case $N(t\send)\ll 1$ is considered
in \cite{p10}. We then have
\be
k_*^2(t\send) = a^2(t\start ) \mu^2 /2 (\mu t\send)\half
. \dlabel{kstarll} \ee
 Using \eqs{classvev}{chik0}  this gives\footnote
{In the prefactor of \eq{classvev}
   we drop a numerical factor and a factor $\tau\mthreehalf$,
because these are negligible compared with the  exponent.}
$\chi\snl^2 \simeq \mu^2 \exp[(4/3) (\mu t\send)\threehalf]$, 
ie.\
\be
\mu t\send
\simeq \( \ln\frac{\chi\snl}\mu \)^{2/3}
. \dlabel{tausvev} \ee
Assuming instead $N(t\send)\gsim 1$, 
\be
k_*^2(t\send) \simeq a^2(t\start )  \mu^{3/2} H^{1/2}
. \ee
Using  \eqs{classvev}{chik0}  we again arrive at 
 \eq{tausvev}\footnote
{Factors $\mu/H$ are  ignored in the prefactor of \eq{classvev}
because the exponent is $\gsim \mu/H$.} 

Our implicit assumption that the growth era starts well before $t\send$
corresponds to $\mu t\send \gg 1$. This is equivalent to $\chi\snl \gg \mu$ or
\be
\min \[ \( \frac H{mgf^{1/5} } \)^{5/3}, \( 10\mfour \frac{fH}{g^4 m} \)^{2/3} \]
\gg 1
. \dlabel{tend1} \ee

The requirement $t\send< t\smax$ corresponds to
\be
\mu t\send < \( \frac m H \)^2 \( \frac H \mu \)^2 = 
\[ 10\mfour \( \frac mH \)^2 \frac fg \]^{2/3}
. \dlabel{tend2} \ee

To get an upper bound on $\mu t\send$ we use $\mu\gsim H$ and 
$\chi\snl^2 \ll  \chivev^2\ll \mpl^2 $, to 
find $\mu t\send \lsim \(\ln (\mpl/H) \)^{2/3}$.
Using the range \eqreff{hbound} this gives an upper bound on
 $\mu t\send$ that 
is of order $5$ to $20$ with the lower end of the range far more likely.
The upper bound corresponds to 
$N(t\send) \lsim  (H/\mu)  (\ln(\mpl/H))\twothird$ giving
\be
N(t\send) \ll  [ \ln(\mpl/H) ]^{2/3} \ll \ln(\mpl/H)
\dlabel{nend} . \ee

%In terms of the parameters $g$, $f$, $m$ and $H$, 
%\be 
%\ln(\chi\snl/\mu) \simeq  \ln \(\frac H{mgf^{1/5}} \) 
%, \dlabel{lnchi} \ee
%and  the requirements for case 1 are
%\bea
%g f^{1/5} &\lsim & H/m\qquad  (\mu t\send \gg1)  \dlabel{case1a} \\
%f &\gsim & 10^3   g (H/m)^2 (\mu t\send)\twothird \qquad  (t\send\lsim t\smax)
%\dlabel{case2a}
%.\eea

Using $\vev{\dot\chi^2(t\start )}\sim \mu^4$ and 
\eqss{sudden}{bhole}{tend1},\footnote
{Only the first case of \eq{tend1} need be invoked for this purpose.} 
we find $\vev{\dot\chi^2(t\start )}\ll \dot\phi^2(t\start)$. This justifies the
assumption  $|\delta t(\bfx,t\send)| \gg |\delta t(\bfx,t\start )|$
made after \eq{deltat3} because the initial strong growth of
$\vev{\dot\chi^2(t)}/\dot\phi^2(t)$ will outweigh the slower variation
of the other factors.

\subsection{The case $t\send > t\smax$}

In this case   
$m^2(t\send)\simeq -m^2$. As we discussed in Section \ref{sszetalin},
$\chi\snl^2$ is given by \eq{chisvev2}, and $\calpzlin$ by \eq{pzetachi1}.
Growth starts, at the latest, at $t\smax+m\mone$.\footnote
{Since growth always begins, $\chi$ always becomes classical. 
We  have no need of a purely
 quantum treatment, which  would require an entirely different 
approach. (In  \cite{p10} we noted $\chi$ may fail to become classical
within the regime $t\ll t\smax$. Contrary to what was stated there,
that does not imply that $\chi$ may fail to become classical at all.)}

Our  our approximation makes $d\mmt/dt$ discontinuous at $t=t\smax$
in violation of the adiabaticity condition \eqreff{adcon2}. In reality
$\mmt$ will be smooth around $t=t\smax$. To avoid specifying a definite
form for $\mmt$, we confine ourselves to the case
$N(t\send)\gsim  1$. Then 
\be
k_*^2(t\send)\simeq a^2(t\start ) m H
. \dlabel{kend}\ee

Assume first that the growth era starts  before $t\smax$. Then 
\eqs{classvev}{chik0} give\footnote
{Since  $mt\send\gg m/H$, and we  ignore
factors of $m/H$ in the prefactor in \eq{classvev}.}
\be
\chivev^2 \simeq  
H^2 \exp \( 2 m(t\send-t\smax ) +  \frac43\( \mu t\smax \)\threehalf \)
. \ee
We also have 
\be
 mt\smax = (\mu t\smax)^{2/3}= (m/\mu)^3 \gg 1
, \dlabel{movermu} \ee
where the inequality holds because we are assuming that growth starts
before $t\smax$. Using the first equality we get
\be
\chivev^2 \simeq
  H^2 \exp \(  2 m \( t\send- \frac13  t\smax) \) \)
\simeq H^2 \exp \(  2 m  t\send \)
. \ee
The final approximation is $t\smax \ll 3 t\send$, which should be 
adequate because we are in the regime $t\smax < t\send$.
In this approximation,  the growth before $t\smax$ has a negligible 
effect. Using it we find 
\be
\chivev^2 \sim H^2 e^{2mt\send}
, \dlabel{chivev2} \ee
 leading to\footnote
{We  ignore a factor $H/m$ within the log, which is permissible
since $H/m$ is also the prefactor.}
\be
N(t\send) \equiv \frac Hm (mt\send) \simeq \frac Hm \ln(\mpl/H)
. \dlabel{ntsvev2} \ee
This gives again the bound  \eqreff{nend}.

Now suppose that growth does not start before $t\smax$.
Then the inequality in \eq{movermu} is reversed leading to 
$N\smax\ll 1$. We therefore arrive again at \eq{chivev2} leading to
\eq{ntsvev2}. In this case $Ht\sub{start}=N\smax +H/m$ which is
$\ll 1$ as before. Also, from 
\eqs{kstart}{kend},  we have $k_*(t\start)/k_*(t\send) \simeq H/m$.
{}Using   \eq{pzetachi9}, this  ensures that the typical
value of $|\delta t(\bfx,t\send)/\delta t(\bfx,t\sub{start})|$
is $\gsim m/H\gg 1$.

\subsection{Duration of the non-linear era}

For case $t\send>t\smax$,  $\chi\send^2$ is the value of $\chi^2$ at which the 
right hand 
side of \eq{fullchi} becomes important, corresponding to
 $\ln(\chivev/\chi\send)\sim 1$.
For the opposite case,  \eqs{chinl2}{bhole}  give 
\be
\frac{\chivev^2}{\chi\send^2} \ll 
\frac{\mpl^2}{mH}
. \dlabel{deltaN2} \ee
This is much less than $\mpl^2/H^2$, which means that 
\eq{deltaN} gives    $N\snl\ll \ln(\mpl/H)$. This  is  
 the same bound that  we obtained for
$N(t\send)$. It therefore applies
to  the total number of $e$-folds of the waterfall,
 $N\sub{water}\equiv N(t\send) + N\snl$.

As seen in Section \ref{seffect}, we need the waterfall to begin more than
22 $e$-folds after the observable universe leaves the horizon,
 if we are to be sure that
 $\calpzlin(k)$ has a negligible effect on cosmological scales.
Equivalently we need $N\sub{obs}- 22 > N\sub{wat}$. 
{}From \eq{nobs} the left hand side of this inequality is bigger than $47-
[\ln (\mpl/H)]/2$ and we have seen that the right hand side is
$\ll \ln(\mpl/H)$. The inequality will therefore hold if 
$47\gg [\ln (\mpl/H)]/2$ ie.\ if  $H/\mpl\gg 10^{-41}$. This is hardly
stronger than the BBN bound \eqreff{hbound}, which means that
$\calpzlin(k)$ is almost certainly negligible on cosmological scales.

\subsection{Two inflation models}

\dlabel{smodels}

To illustrate the power of our results we apply them to two inflation 
models based on supersymmetry.

\subsubsection{Supersymmetric GUT hybrid inflation}

Supersymmetric GUT hybrid inflation \cite{gut1,gut2}
takes $\chi$ to be a GUT Higgs field
so that $\chivev\simeq 10\mtwo\mpl$ corresponding to $(H/m)^2\simeq 10\mfive$.
This is not small enough for the  `end of inflation' formula to yield
the entire waterfall contribution to $\zeta$ (\eq{criterion2}).
Supersymmetry gives  $g^2=2\lambda$ leading to  $g^2 = 10^9 (H/\mpl)^2$.
This leaves for our discussion two independent  parameters which we take
as $g$ and $f$.

The potential $V(\phi)$ may depend on several parameters \cite{gut2}.
It typically steepens, and our  discussion applies
only if the parameters are such that steepening does not 
end slow-roll before $t\send$. 
Requiring the  inflaton perturbation to generate $\zeta$ on cosmological
scales, the steepening implies $10^{-1.5} g \lsim f\lsim 1$, the lower bound
coming from \eq{fgg}.
Using \eq{sudden} we have $\mu/H \simeq 10^2 (g/f)\third$.
The fast transition requirement \eqreff{sudden} is  certainly satisfied 
if  $g^2 \gg  10^{-12}$ (ie.\ $H/\mpl \gg 10^{-10}$
which  usually holds \cite{gut2}.

The parameter space allows $t\send<t\smax$ (with either of the
possibilities in \eq{chisvev1}) as well as $t\send>t\smax$.
Provided that  $H/\mu$ is well below 1, the 
 duration of the waterfall is quite short, and the `end of inflation'
formula can give $\zeta\snl$ 
 in part of the parameter space (\eqreff{criterion}).

\subsubsection{Supernatural/running-mass inflation}

Supernatural inflation \cite{supernatural} and running-mass 
\cite{running} inflation take 
 $\chivev$ roughly of order $\mpl$ corresponding to $m$ roughly of order
$H$. This can be motivated by supposing that $\chi$ is a string modulus,
with gravity-mediated or anomaly-mediated supersymmetry breaking
\cite{stringaxion}. The  former case, 
$V_0\quarter\sim 10^{10}\GeV$ or $H\sim 10^{-15}\mpl$ is usually invoked and
the latter would give $H\sim 10^{-13}\mpl$ or so. This  low inflation scale
and $m\sim H$ are distinguishing features of the  paradigm.

The potential for supernatural inflation is $V(\phi)=m_\phi^2 \phi^2/2$
which does not allow $\zeta=\zeta_\phi$ on cosmological scales.
Running-mass inflation takes $V(\phi)$ to be the renormalization group
improved potential allowing $\zeta=\zeta_\phi$ on cosmological scales
which is assumed. In a suitable regime of parameter space, 
$\zeta_\phi(k)$ on small scales can be big enough to exceed the 
cosmological bound on black hole formation, providing a constraint on
the parameter space; in  other words we can have $f\sim 10^3$.
This is another distinguishing feature of the paradigm.

Since $m$ is roughly of order $H$ our criterion $m^2\gg H^2$ cannot be very well satisfied
and the analysis of the next Section is really more appropriate. To proceed we 
assume that  $m/H$ is a bit above 1, and take   $f\sim 1$.
Then the fast transition requirement $H\ll \mu$ is satsfied for 
$g^2\gg 10^{-8}$ which is expected.

Since $m$ is roughly of order $H$ and we deal with the case
 $t\send > t\smax$,
corresponding to $m^2(t\send)\simeq -m^2$. This gives
$N(t\send) \sim \ln (\mpl/H) \sim 33$,  and $k_*(t\send)$ is outside the horizon, with
$\calpzlin(k_*(t\send)) \simeq (H/2m)^2$. This is is not far below 1, and 
the black hole bound might be violated.

The duration of the non-linear era is 
 $N\snl \sim 1$. Since \eq{criterion} is not satisfied, the
 contribution  $\zeta\snl$ is not given by the `end  of inflation' formula.
 
\section{The case $m\sim H$}

\dlabel{smconst}

 Now we consider the 
case that  $m/H$ is $\lsim 1$ but not extremely small.
 To arrive at estimates we   assume  that 
$\calp_\chi(k)$  in this regime
continues to peak at some value $k_*(t) \ll a(t)\mmt$.
Since $\mmt\leq m\sim H$ this means that $k_*(t)$ is always outside the
horizon.

We assume that the gradient of $\chi$ is negligible, 
checking the self-consistency
of that assumption later. 
Then \eq{chiddot2} holds. Considering 
 either of the two independent solutions we define
$s(t)$ by
\be
\dot\chi(\bfx,t)= H s(t) \chi(\bfx,t)
, \dlabel{sdef} \ee
giving 
\be
s^2(t)+ 3s(t) -  \mmts/H^2 = - \dot s/H 
. \dlabel{seq} \ee

We  assume that the right hand side  
of \eq{seq} is negligible, checking later for
self-consistency.
%\be
%s^2(t)+ 3s(t) = \mmts/H^2 
%. \dlabel{ssq} \ee
Keeping only the  growing mode this gives
\bea
s(t) &=& - \frac32 + \sqrt{\frac94 + \frac{\mmts}{H^2} }  \dlabel{sofbeta} \\
&\simeq&  \frac13 \frac {\mmts}{H^2}   \dlabel{sofbeta2}
. \eea
The gradient of $\chi$ is indeed negligible compared with
$\dot \chi$, which means that $\rho_\chi$ and $p_\chi$ are given by 
\eqs{rchi2a}{pchi2a}. Also, the assumption that the right hand side of \eq{sofbeta}
is negligible is self-consistent if and only if
 \eq{condition} holds. This means that the field equation \eqreff{chiddot2} is
consistent with the energy continuity equation.

If  the approximation \eqreff{approx1} holds, \eq{condition} becomes
$Ht \gg 1$. But the same approximation inserted into \eq{sdef} gives at
$Ht \gg 1$
\be
Ht \sim (H/\mu)^2 \ln \[ \chi^2(\bfx,t)/\chi^2(\bfx,H\mone) \]
. \dlabel{htsim} \ee
Since we haven't calculated $\chi(\bfx,t)$ from the vacuum fluctuation we 
don't know the precise  value of $\chi^2(\bfx,H\mone)$ but 
  it  presumably lies roughly between $\mu$ and $m\sim H$ since these are the
relevant mass scales. As we saw earlier this would make the log at most of order
$10^2$ or so. Therefore, since we are imposing $H\ll \mu$, \eq{htsim} is hardly
compatible with $Ht \gg 1$. The conclusion is that \eq{condition} probably requires
the regime \eqreff{approx2}, $m^2(t) \simeq -m^2$, which we assume from now on. 
That in  turn implies
$\vev{\dot\chi^2(t\send)}\gg  \dot\phi^2(t\send) $. 

To calculate $\zeta\slin$ we use \eq{zchi}, and assume 
$|\delta t(t\send)/\delta t(t\sub{start})| \gg 1$.
Using \eqss{rchi2a}{sofbeta}{sdef}, we find 
\be
\rho_\chi/\dot\chi^2=3/2s(t) \simeq H^2/m^2
, \ee
to be compared with
$\rho_\chi/\dot\chi^2 =3H/2\mmt$ in the case $m^2\gg H^2$.  We therefore have
\be
\zeta\slin(\bfx) \simeq  -\frac {H^2}{m^2} 
 \frac{
\delta \chi^2(\bfx,t\send)
}{
\vev{\chi^2(t\send)} 
} \dlabel{zetachi2}
. \ee 

Since $\calp_\chi(k)$ peaks at $k_*$, we expect that the final equality of 
\eq{classvev} will be roughly correct. Also we expect that 
$\calp_{\delta \chi^2}(k)$ will be given roughly by \eq{pchisq1} 
 at $k\lsim k_*$ and will fall off at bigger $k$. Then, using \eq{sofbeta2}
 we see that  
$\calp_{\zeta\slin}(k) $  peaks  at $k\sim k_*(t\send)$ with a  value
\be
\calp_{\zeta\slin}(k_*(t\send)) \simeq \frac{H^2}{m^2} 
. \ee
We conclude that  the black hole bound is likely to be violated if $m$
is significantly below $H$.
 
A crucial feature of our setup is the
condition \eqreff{condition},  which is necessary for consistency 
if the  gradient of $\chi$ is negligible and  there is no cancellation between the
 two terms of $\rho_\chi$. We found that the solution of \eq{fulleq} then indeed
makes the gradient of
$\chi$  negligible, with no cancellation. But the 
 solution of \eq{fulleq} may also make  the 
gradient negligible with no cancellation, 
in a part of parameter
space where  with the condition  \eqreff{condition} violated. 
In such a regime,  we would have
 to conclude that the linear approximation leading to
\eq{fulleq} is invalid.

\section{Comparison with other calculations}

Nineteen other  papers have 
 considered  the contribution of the waterfall to $\chi$
\cite{supernatural,glw,bfl,bdd,msw,af,afs,ptba,tpb,afn,fsw,gs,ev,bc1,bc2,hc,ejmmv,bk,tsy,bk2}.
in the  fast transition regime.
Some   of them also 
consider the issue of black hole formation \cite{supernatural,glw,bk,bk2}, concluding that 
the black hole constraint is satisfied for
 $m\gg H$ but not for $m\sim H$. That is roughly our conclusion though we are less
sure. In view of this, one  may  wonder whether the 
 present paper and its companion  \cite{p10} are needed. They are, for several  reasons.

First,
 all of the previous papers take $\phi$
to be canonically normalized, and nearly all of them 
go much further by assuming   $V(\phi)=m_\phi^2\phi^2/2$. 
Second,   none of the previous papers specifies all of the assumptions that are made,
as we do here and in \cite{p10}.
Third,  none of them except \cite{bk,bk2} considers
the non-gaussian black hole bound as we have done in the present paper. 
Fourth,  most of them present a calculation which is much more complicated than ours. 
Finally, all of the other papers  except perhaps \cite{bk,bk2} have errors. 

The last point was considered in our earlier paper \cite{p10}. 
The problem for many of  the papers  \cite{supernatural,glw,bfl,bdd,msw,af,afs,ptba,tpb}
is that the waterfall is treated as two-field
inflation, without imposing the requirement
$\vev{\bar\chi^2}\gg \vev{\chi_>^2}$ that would be needed to 
justify such a treatment.\footnote
{The papers \cite{glw,af,supernatural,afn} set  $\bar\chi^2 = \vev{\chi^2}$
 while the others regard it as
a free parameter.}
As we have seen, this is not  the case.
It is only within the slow transition regime,  considered in  
 \cite{tsy,clesse,kkn,mor}, 
that one can expect to find a regime of parameter space that allows
 the waterfall to be treated as two-field inflation.

 We will not repeat the analysis of the problems of the
other papers \cite{fsw,gs,ev,bc1,bc2,hc,ejmmv}  appearing before \cite{p10}, that was given
in the latter paper.
After \cite{p10} three   papers appeared \cite{afs,bk,bk2}. The paper  
 \cite{afs}  is  a continuation of \cite{gs}.  Its
 main focus is on the case $\phi\sub c\gsim 10\mpl$,
where inflation continues after the waterfall, but that  does not affect the
contribution to $\zeta_\chi$ generated during the waterfall, and 
Eq.~(6.15)) of \cite{afs}  reproduces the expression  for $\zeta_\chi$ given in 
Eq.\ A(10) of \cite{gs}. 

The papers \cite{bk,bk2} consider the case
 $m\lsim H$, and they calculate $\chi_k$ by numerical integration with the potential
$V(\phi)=m_\phi^2\phi^2/2$.\footnote
{With $m\lsim H$, the fast transition requirement
\eqreff{fastquad} conflicts with  the slow-roll requirement
$m_\phi\ll H$, but the two are roughly compatible with the 
choice $(m_\phi/H)^2=10\mone$ of  \cite{bk,bk2}.}
Then they evaluate $\zeta\slin$ by integrating \eq{zetadot}, 
finding that  $m<H$ is definitely forbidden by the black hole bound.
The calculation assumes that the gradient of $\chi$ can be ignored when evaluation
$\rho_\chi$ and $p_\chi$ but they don't investigate the compatibility 
of the evolution equation
\eq{fulleq} with the energy continuity equation. 
However, their results for the case
$m=H$ (with the other parameters fixed at particular values)
shown in their  Figure 2 is in excellent agreement with our \eq{sdef},
assuming    $\mmt=m$.  Their result for
 $\calp_{\zeta\slin}(k_*(t\send),t\send)$ with the same parameter choice, 
shown in  their Figure 4,  is also in agreement with ours, assuming   in addition
 $\vev{\dot\chi^2(t\send)}\gg \dot\phi^2(t\send)$. 
It therefore seems that for at least this parameter choice,  
their assumption that the 
 gradient of $\chi$ is negligible is justified, and that moreover the consistency
condition \eqreff{condition} is satisfied. 
But there is no reason to think that
the same is true in the entire parameter space,  considered in their Figure 5.
Regarding the black hole bound, they note that $\zeta\slin$ has the non-gaussian
form \eqref{zetang}. In \cite{bk} they use $\vev{\zeta\slin^2}\lsim 1$
instead of our $\calpzlin\lsim 1$. As the
 width of the peak in $\calpzlin(k)$ is rather broad, this will somewhat  overestimate
the region of parameter space forbidden by 
 the cosmological bound on $\calpzlin$ as we noted in Section \ref{sbh}.
In \cite{bk2} a more sophisticated procedure is used to obtain the black hole bound,
but they don't estimate the theoretical error and it is unclear to us whether it
represents an improvement on our rough estimate $\calpz\lsim 1$.

\section{Conclusion}

\dlabel{sconc}

We have considered  the contribution  $\zeta\slin$ to $\zeta$, that is 
generated during the 
linear era of the waterfall within the Standard Scenario. 
We gave a rather complete calculation for the case that the waterfall mass $m$
is much bigger than $H$, and  arrived at estimates for $m\sim H$.

Taking on board our discussion of the 
non-gaussian black hole bound, we concluded that 
the black hole bound will be satisfied for $m\gg H$, but that it may well be violated
 for $m\lsim H$.
The latter case will be further investigated in a future publication \cite{next},
by numerically integrating \eq{fulleq}.

A lot more will have to be done before we 
have a complete understanding of contribution to 
$\zeta$ generated during the waterfall. A fundamental problem is to handle the 
ultra-violet cutoff, that is needed to obtain finite values for the fields and
for the energy density and pressure. Our procedure of keeping only the classical
field modes is approximate,  and it violates at some level the energy continuity
equation. This and related issues are discussed for instance in \cite{maggiore}.
A precise procedure is advocated in \cite{laura}, but its
 relation to our approximate procedure is unclear.

An understanding of the ultra-violet cutoff will allow one to decide on the 
minimum value of $\dot\phi(0)$ that allows an initial linear era \cite{dufaux,p10}.
With that in place one would hopefully verify that the value invoked in the present
calculation is big enough.
But it will still be unclear how to evaluate the contribution to $\zeta$
that is generated during inflation after the linear era ends, when it is not
given by the `end of inflation' contribution. A numerical simulation, even with
reasonable simplifications, might well require one to consider a patch of the 
universe that is too big to handle.

%%%%%%%%%%%%%%%%%%%%%%%%%%%%%%%%%%%%%%%%%%%%%%%%%%%%%%%%%%%%%%%%%%%%%%
\section{Acknowledgments}
%%%%%%%%%%%%%%%%%%%%%%%%%%%%%%%%%%%%%%%%%%%%%%%%%%%%%%%%%%%%%%%%%%%%%%
The author acknowledges support from the Lancaster-Manchester-Sheffield Consortium for
Fundamental Physics under STFC grant ST/J00418/1, and from
 UNILHC23792, European Research and Training Network (RTN) grant. 

%%%%%%%%%%%%%%%%%%%%%%%%%%%%%%%%%%%%%%%%%%%%%%%%%%%%%%%%%%%%%%%%%%%%%%

\end{document}